\documentclass[namedreferences]{solarphysics}

\usepackage[optionalrh]{spr-sola-addons} 
\usepackage{graphicx}        
\usepackage{natbib}         

\usepackage{amsmath}

\renewcommand{\leq}{\leqslant}  
  
\begin{document}


\begin{article}

\begin{opening}

\title{Multiple Scattering of Seismic Waves from Ensembles of Upwardly Lossy Thin Flux Tubes}

\author[addressref={aff1},email={email:hanson@mps.mpg.de}]{\inits{C.S.}\fnm{Chris S.}~\lnm{Hanson}}
\author[addressref={aff1,aff2},corref,email={email:paul.cally@monash.edu}]{\inits{P.S.}\fnm{Paul S.}~\lnm{Cally}}
       
\runningauthor{C.S. Hanson and P.S. Cally}
\runningtitle{Multiple Scattering from Upwardly Lossy Flux Tubes}

\address[id=aff1]{Max-Planck-Institut f\"ur Sonnensystemforschung, Justus-von-Liebig-Weg 3, 37077 G\"ottingen, Germany}
\address[id=aff2]{School of Mathematical Sciences and Monash Centre for Astrophysics, Monash University, Melbourne, Victoria 3800, Australia}


\begin{abstract}
Our previous semi-analytic treatment of $f$- and $p$-mode multiple scattering from ensembles of thin flux tubes  (Hanson and Cally, \emph{Astrophys.~J.} \textbf{781}, 125; \textbf{791}, 129, 2014) is extended by allowing both sausage and kink waves to freely escape at the top of the model using a radiative boundary condition there. As expected, this additional avenue of escape, supplementing downward loss into the deep solar interior, results in substantially greater absorption of incident $f$- and $p$-modes. However, less intuitively, it also yields mildly to substantially smaller phase shifts in waves emerging from the ensemble. This may have implications for the interpretation of seismic data for solar plage regions, and in particular their small measured phase shifts.

\end{abstract}

\end{opening}


\section{Introduction}The interaction of the Sun's $f$- and $p$-mode oscillations with large collections of thin magnetic-flux tubes that make up plage and network is important for our understanding of the broader seismology of solar activity. Individual tubes absorb, phase shift, and scatter incident waves, but close assemblies of tubes provide a multiple-scattering regime and collective behaviour that makes them more than just a summation of their parts.

The historical context is addressed in detail by \citet{HanCal14aa,HanCal14ab}. Briefly, early modelling of multiple scattering in ensembles of flux tubes ignored gravitational stratification due to the considerable mathematical complications that arise \citep{BogZwe85aa,ZweBog86aa,BogZwe87aa,RyuPri93aa,RyuPri93ab}. 

Techniques for dealing with scattering from single tubes in stratified atmospheres were developed by, for example, \citet{HanBirBog08aa} and \citet{HinJai12aa}. \citet{JaiHinBra09aa}, \citet{JaiGasHin11aa}, and \citet{JaiGasHin11ab} considered multiple tubes, but did not allow for multiple scattering, thereby treating the ensembles as simply the sum of the individual tubes. The true multiple-scattering regime was attacked by \citet{HanCal09ab} for the case of two adjacent tubes, adapting the formalism of \citet{kagyue86aa} from the engineering literature. Only kink waves were modelled, as it was not then clear how to implement the proper side boundary conditions for sausage modes. The role of the acoustic jacket \citep{BogCal95aa,Cal13aa} was emphasised in mediating near-field coupling between tubes. 

The proper approach to side boundary conditions for sausage modes was resolved by \citet{AndCal11aa} and \citet{HinJai12aa}, and implemented by \citet{HanCal14aa}  for two tubes, and \citet{HanCal14ab} for multiple tubes, again using the \citet{kagyue86aa} formalism. A slight modification was made to correct an unphysical symmetry breaking that appeared in the results of \citet{HanCal09ab}. Direct numerical simulations (not using the thin-tube approximation) yielded general confirmation of the results \citep{FelCroBir13aa,Dai14aa}.

Wave absorption and phase advancement have been measured in solar plage \citep{bra95aa,ChoSunCha00aa}. \citeauthor{bra95aa} reported no measurable phase change in waves emerging from plage, but \citeauthor{ChoSunCha00aa} found a weak positive shift. On the other hand, phase shifts calculated for thin-tube ensembles by \citet{HanCal14ab} are always negative. The likely cause of this discrepancy is that the ensemble models postulate field-free quiet-Sun plasma between the flux tubes, whereas real plage is both thermally different from quiet Sun and probably has a continuum of weak field in addition to the many ``discrete'' flux tubes. Clearly, continuum field speeds up transiting waves (which are fast MHD waves in magnetic environments) thereby advancing their phase, whereas multiply scattering discrete ensembles slow them down in the manner of a ball in a pin-ball machine, resulting in negative phase shifts. The competition between these two effects may in part be responsible for the small measured shifts from plage.

To date, it has been customary to impose a stress-free boundary condition at the top of each tube, thereby precluding any upward wave losses. The purpose of this article is to investigate the implications of relaxing that restraint.


\section{Upwardly Radiative Boundary Conditions}
The mathematical formalism developed in \citet{HanCal14aa,HanCal14ab} addresses the absorption and scattering of waves by single and multiple flux tubes. The ``absorbed'' waves take the form of sausage or kink tube waves, with their attendant acoustic jackets, that propagate freely down into the solar interior and are therefore lost to the observable surface oscillations. This is the nature of absorption in those models. However, in reality, waves may also propagate upward into the solar atmosphere. This was expressly prohibited in our previous work by imposing perfectly reflective stress-free boundary conditions at the top of each tube, following \citet{BogHinCal96aa}. 

There have been several attempts to relax this restriction. \citet{CroCal99aa} allowed the thin tube to continue upward through a crude model chromosphere and corona without expanding in radius and matching onto outgoing magneto-acoustic waves in the corona. This of course is also very unrealistic, since the tube will inevitably expand well beyond the thin-tube regime very quickly. On the other hand, \citet{HinJai08aa} sidestep the issue of a realistic boundary condition by introducing the concept of a maximal-flux condition. This is applied in detail to sausage modes in single flux tubes by \citet{GasJaiHin14aa}.

Adopting the truncated adiabatic polytrope external model of \citet{BogHinCal96aa}, with polytropic index $m_p$, truncation depth $z_0$, dimensionless depth variable $s=-z/z_0$ ($1\leq s<\infty$), and assuming the thin flux tubes are of uniform plasma $\beta$ (the ratio of gas to magnetic pressure), the thin tube oscillation equation for the sausage mode (azimuthal order $m=0$) is  
\begin{multline}
\left[ \omega^2(2m_p+\beta(m_p+1)) + \frac{2gs}{z_0}\frac{\partial^2}{\partial s^2} + \frac{g(m_p+1)}{z_0}\frac{\partial}{\partial s} \right]\xi_{\|}  \\ = -\omega^2(m_p+1)(\beta+1)\frac{\partial \Psi_{\rm inc}}{\partial s} 
\end{multline}
and that of the kink mode ($m=\pm1$) is
\begin{multline}
\left[ \omega^2 z_0 +\frac{2gs}{(1+2\beta)(m_p+1)} \frac{\partial^2}{\partial s^2}+\frac{g}{1+2\beta} \frac{\partial}{\partial s} \right]\xi_{\perp}
=
\frac{2(1+\beta)}{1+2\beta}\omega^2z_0\frac{\partial \Psi_{\rm inc}}{\partial x}. 
\end{multline}
Here $\xi_{\|}$ and $\xi_\perp$ are respectively the longitudinal and transverse plasma displacements, $g$ is the gravitational acceleration, and $\omega$ is the circular frequency. The inhomogeneous term $\Psi_{\rm inc}$ represents the external driving by $f$- and $p$-modes. Polytropic index $m_{\rm p}=3/2$, corresponding to adiabatic index $\gamma=5/3$, is used throughout.

These driven tube equations may be solved subject to appropriate boundary conditions at the top ($s=1$) and bottom ($s=\infty$) by constructing Green's functions. As usual, the Green's function is built from the homogeneous solutions satisfying suitable homogeneous boundary conditions. These may be expressed in terms of Hankel functions \citep{CroCal99aa},
\begin{gather}
\psi_\sigma(s) = s^{-\mu/2} H_\mu^{(1)}\left(2\nu\sqrt{\epsilon_\sigma s}\right), \\[4pt]
\phi_\sigma(s) = s^{-\mu/2} \left[H_\mu^{(2)}\left(2\nu\sqrt{\epsilon_\sigma s}\right) + \lambda_\sigma
H_\mu^{(1)}\left(2\nu\sqrt{\epsilon_\sigma s}\right)\right],
\end{gather}
where $\mu=(m_p-1)/2$, $\nu^2=m_p\omega^2z_0/g$, $\sigma$ represents either $\|$ (sausage) or $\perp$ (kink), and we have adopted the convenient parameters
\begin{equation}
\epsilon_{\|}=\frac{2m_p+\beta(m_p+1)}{2m_p} \mbox{ and } \epsilon_\perp = \frac{(m_p+1)(1+2\beta)}{2m_p}.
\end{equation}

The solution $\psi_\sigma$ is constructed to satisfy a pure radiation condition at $s=\infty$, based on the well-known large-argument asymptotic behaviour of the Hankel function. The parameter $\lambda_\sigma$ may be chosen to select an appropriate boundary condition at the top [$s=1$]. For example, a particular choice (that is not used here) recovers the stress-free solution in Equation (4.4) of \citet{BogHinCal96aa}. However, here
we choose simply $\lambda_\sigma=0$, thereby imposing an outgoing radiation condition at $s=1$ (see also Equation~(4.7) of \opencite{HinJai08aa}). The identification of the $H_\mu^{(2)}$ Hankel function as ``purely upgoing'' is not trivial, as the exponential asymptotic behaviour does not (generally) apply there. Nevertheless, \citet{Cal12aa} has shown that the Hankel function at small argument does actually represent a unidirectional wave coupled with an unavoidable wake (unless $\mu$ is half an odd integer) that decays algebraically in time. On that basis, we identify the $H_\mu^{(2)}$ solution as the radiation solution at the top. 

The rest of the analysis is exactly as set out by \citet{HanCal14aa,HanCal14ab}.


\begin{figure*}
\begin{center}
\includegraphics[width=\hsize
]{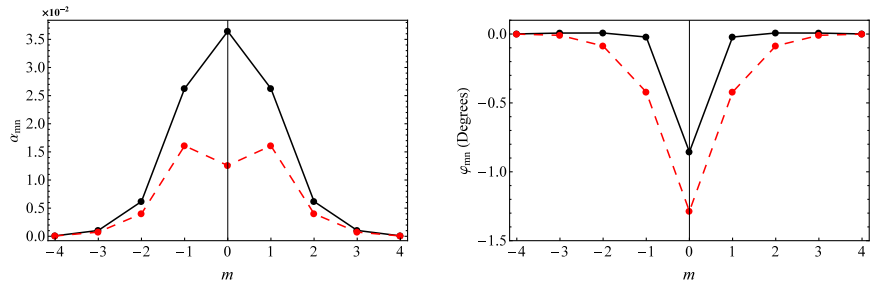}
\caption{The absorption and phase shift for 3 mHz $f$-modes ($n=0$) of various cylindrical orders $m$ incident upon two tubes of $\beta=1$ separated by $0.25 \lambda$ (1.225 Mm) positioned along the $x$-axis. The two cases studied here are the stress-free (dashed red) and the radiative (full black) boundaries.}
\label{fig:2tube}
\end{center}
\end{figure*}

\section{Results}
For conciseness, we will mainly consider incident $f$-modes, as these interact most strongly. Corresponding results for the $p_1$ mode will be presented very briefly.

First, we address the case of two $\beta$ =1 flux tubes (one at coordinate centre) separated by a quarter of an incident $f$-mode wavelength ($0.25\lambda$). The absorption and phase shift for the two top boundary conditions, stress free and radiation, are compared in Figure~\ref{fig:2tube}. Absorption $\alpha_{mn}$ and phase shift $\varphi_{mn}$ of an incident wave of cylindrical degree $m$ and radial order $n$ are defined by \citet[Equations (4) and (5)]{HanCal14ab}. The $f$-mode corresponds to $n=0$. Not surprisingly, the ability of the tubes to lose energy upward roughly doubles the absorption coefficients for both the sausage ($m=0$) and kink ($m=1$) modes. Higher-order modes ($|m|>1$) are weakly seen by the system as the off-axis tube reacts to $m=0$ and 1 components in the incident wave as experienced from its own coordinate centre. On the other hand, the phase shift is substantially reduced by the open boundary condition, with only the sausage mode retaining any significant shift.

\begin{figure*}
\begin{center}
\includegraphics[width=\hsize,clip=true,trim=110 0 110 0]{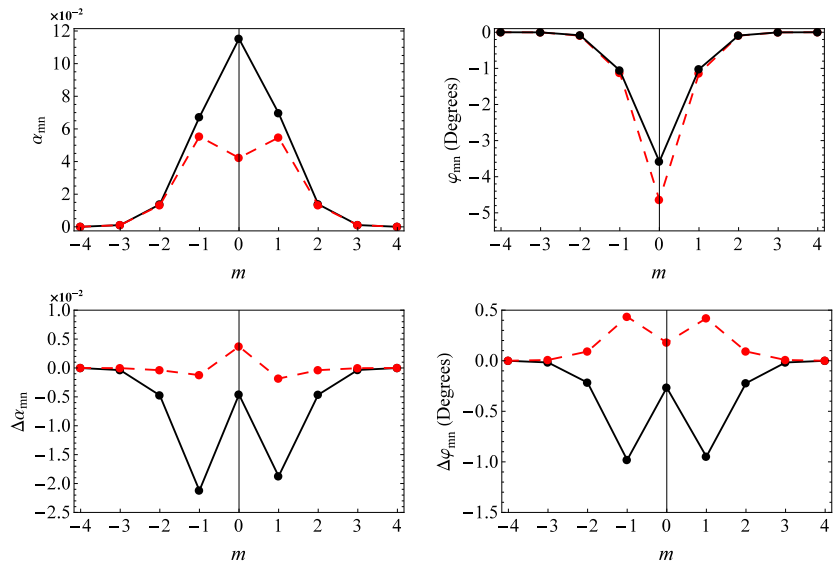}
\caption{Top: the absorption and phase shift (degrees) of 3 mHz $f$-modes ($n=0$) of cylindrical order $m$ that are incident upon an ensemble of seven identical tubes ($\beta =1$) randomly situated around the origin (as in \opencite{HanCal14ab}).  The dashed red curve shows the scattering properties of the stress-free boundary condition and the full black the transmissive boundary condition. The bottom two frames show the change in absorption and phase from the case of the non-interacting ensemble (multiple scattering $-$ single scattering). The random positioning of the tubes breaks $m\to-m$ symmetry, as is most apparent in the bottom-left panel.}
\label{fig:7tubediff}
\end{center}
\end{figure*}

Figure \ref{fig:7tubediff} displays similar results for the same randomly positioned seven-tube ensemble of $\beta=1$ tubes as described in Table 2 of \citet{HanCal14ab}. Once again there is a clear difference between the stress-free and radiative top boundary cases, with absorption of the sausage mode more than doubled and the negative phase difference slightly reduced in magnitude. The lower panels illustrate the importance of the multiple scattering regime. Multiple scattering has a far larger effect for kink modes ($|m|=1$) in particular where the radiative top boundary is applied.

\begin{figure}
\begin{center}
\includegraphics[width=\hsize]{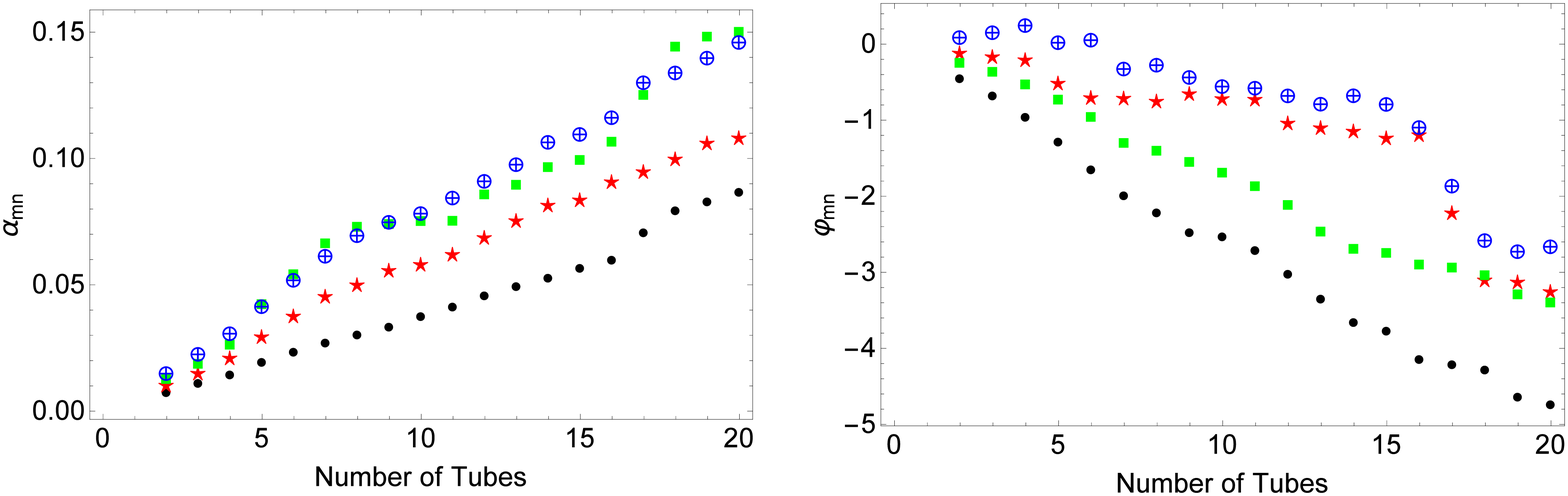}
\caption{The absorption (left) and phase shift (degrees, right) of $f$-modes incident on a randomly distributed ensemble of tubes as a function of increasing tube number. The tubes are identical ($\beta =1$) and can scatter up to $|m|=4$. The stress-free boundary cases are shown by the black dots and red stars (identical to \opencite{HanCal14ab}) for the $m=1$ and $m=0$ scatter, respectively. The radiative tubes are shown by the green squares and blue circled crosses for the $m=1$ and $m=0$ waves, respectively.}
\label{fig:tubenumber}
\end{center}
\end{figure}

Figure \ref{fig:tubenumber} again extends some results from \citet{HanCal14ab}. Absorption and phase shift are plotted for the sausage and kink modes, with either the stress-free or radiative boundary condition, for between 2 and 20 identical randomly placed $\beta=1$ tubes as they are added, one after the other, within a radius of 2.5 Mm from the origin. The precise details of the results depend of course on the random placement, but broadly absorption is significantly enhanced in all cases by opening up the top boundary, and increases roughly linearly with tube number. The phase shift is reduced in magnitude strongly for the kink wave and slightly for the sausage wave.

\begin{figure}
\begin{center}
\includegraphics[width=\hsize]{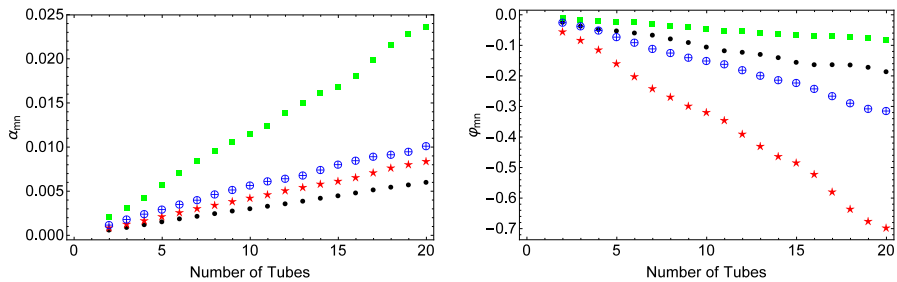}
\caption{The absorption (left) and phase shift (degrees, right) of $p_1$-modes incident on a randomly distributed ensemble of tubes as a function of increasing tube number. The tubes are identical ($\beta =1$) and can scatter up to $|m|=4$. The stress-free boundary cases are shown by the black dots and red stars (identical to \opencite{HanCal14ab}) for the $m=1$ and $m=0$ scatter, respectively. The radiative tubes are shown by the green squares and blue circled crosses for the $m=1$ and $m=0$ waves, respectively.}
\label{fig:tubenumber_p1}
\end{center}
\end{figure}

Figure \ref{fig:tubenumber_p1} repeats these calculations, but for incident $p_1$- instead of $f$-modes. For each boundary condition, both the absorption and phase shift are reduced by roughly an order of magnitude compared to the $f$-mode case. Once again, net absorption is increased by opening the flux-tube tops to wave egress, especially for the kink mode. The phase shifts roughly halve in both cases. 

\begin{figure}
\begin{center}
\includegraphics[width=\hsize]{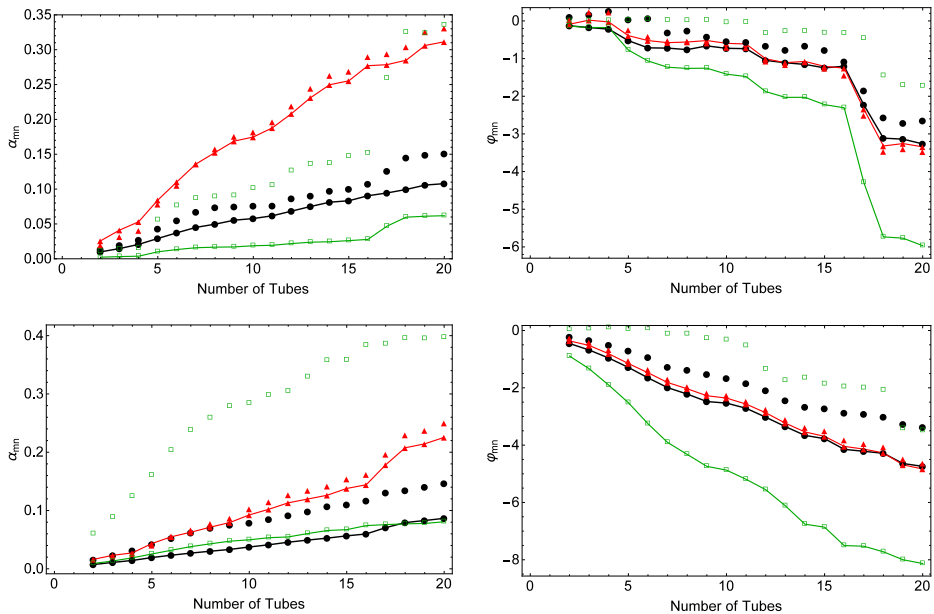}
\caption{The absorption (left) and phase shift (degrees, right) of the $m=0$ (top) and $m=1$ (bottom) incident and outgoing $f$-mode waves interacting with a random ensemble of growing tube number of identical properties ($\beta = 1$: filled black circles; $\beta = 0.1$: filled red triangles; $\beta = 10$: open green squares) with the radiative (points with lines) and stress free (points only) boundary conditions.}
\label{fig:BETA}
\end{center}
\end{figure}

Finally, in Figure \ref{fig:BETA} we again compare the 2--20 randomly placed tubes, but for $\beta=1$, 0.1, and 10. Absorption of sausage modes (top-left panel) is most pronounced for the strongest tubes ($\beta=0.1$), though note that through happenstance, tubes 17 and 18 produce a particularly strong increase in absorption for $\beta=0.1$. The phase shift is broadly reduced by the radiative boundary, especially for the weak tubes $\beta=10$.

For the kink modes (bottom panels of Figure \ref{fig:BETA}), the weak-tube radiative case is clearly dominant in absorption, and most muted in phase shift. This compares with the stress-free boundary case, where $\beta=10$ has the greatest phase shift.

\section{Conclusions}

Broadly, allowing tube waves to escape at the top into a notional solar atmosphere increases the total absorption of an ensemble of thin flux tubes, as we might expect. Details vary, depending on plasma $\beta$, tube arrangement, and mode (sausage or kink, $f$ or $p_1$), but the trend is consistent. The most spectacular increase is seen for both $m=0$ and $m=1$ $f$-modes incident on a collection of up to 20 weak tubes ($\beta=10$), with strong tubes' absorption increased only marginally.

The more surprising result though is that the wave losses at the top of the tubes generally reduce the overall phase shift produced by the ensemble, particularly for weak tubes. This applies to both $f$- and $p_1$-modes. We do not have a simple physical explanation for this, but it is significant for the interpretation of the seismology of plage, and potentially provides a means of estimating the strengths of constituent flux tubes on average.

The greatly reduced absorption of $p$-modes relative to $f$-modes was documented in Figure 5 of \cite{HanCal14ab} for the stress-free top boundary, and carries over to the radiative case. Nevertheless, the seemingly linear increase in absorption with number of tubes suggests that even $p$-modes could measurably suffer absorption in plage, especially if they have lossy tops. This reduced effect with increasing radial order is strong evidence that the scattering and absorption are closely tied to the surface. 

Of course, in reality the behaviour of tube waves when they reach the photosphere will not be as simple as suggested by the radiative boundary condition. It will depend on the rapid flux tube expansion expected there, field connectivity, frequency, and many other details of the physics. Our purpose here has merely been to explore the consequences of the most extreme behaviour, total loss to the atmosphere of tube waves reaching the surface. This gives some insight into the range of behaviours that we might expect to see in the real Sun.

\begin{acks}[Acknowledgements]
Paul Cally would like to acknowledge the kind hospitality of Laurent Gizon and the Max Planck Institute of Solar System Research in G\"ottingen during his visit in March--April 2015, when this work was done.
\end{acks}

\section*{Disclosure of Potential Conflicts of Interest}
The authors declare that they have no conflicts of interest.



\bibliographystyle{spr-mp-sola}        
\bibliography{fred}

\end{article}
\end{document}